\documentstyle[epsfig,eqsecnum,aps]{revtex}
\begin{document}
\title{Gamow-Teller strength distributions in Fe and Ni stable isotopes}
\author{P. Sarriguren, E. Moya de Guerra, and R. \'Alvarez-Rodr\'{\i}guez}
\address{Instituto de Estructura de la Materia,\\
Consejo Superior de Investigaciones Cient\'{\i }ficas,\\ 
Serrano 123, E-28006 Madrid, Spain}
\maketitle

\begin{abstract}
% Text of abstract
We study Gamow-Teller strength distributions in some selected nuclei 
of particular Astrophysical interest within the iron mass region.
The theoretical framework is based on a proton-neutron Quasiparticle 
Random Phase Approximation built on a deformed selfconsistent mean 
field basis obtained from two-body density-dependent Skyrme forces. 
We compare our results to available experimental information obtained 
from $(n,p)$ and $(p,n)$ charge exchange reactions.
\end{abstract}


\section{Introduction}

It is well known \cite{bethe,ffn} that nuclear $\beta$-decay and electron 
capture processes are very important mechanisms to understand the late 
stages of stellar evolution. These processes are essential ingredients in 
calculations of supernova formation. In particular, Gamow-Teller (GT) 
properties of nuclei in the region of medium masses around A=56 are of 
special importance because they are the main constituents of the stellar 
core in presupernovae formations.

It is clear that due to the extreme conditions of densities and temperatures
that hold in the stellar scenarios, involving highly unstable nuclei as 
well, most of those properties cannot be measured directly. Therefore, the 
GT strength distributions must be estimated in many cases by 
model calculations. 
Collapse simulations of supernovae have been carried out so far by 
treating the GT transition rates in a rather qualitative way. For 
example, it is a common practice \cite{ffn} to assume that the whole GT 
strength resides in a single resonance whose energy relative to the 
daughter ground state is parametrized phenomenologically, taking the total 
GT strength from the single-particle model. 

In the last decades, $GT_+$ strength distributions on nuclei in 
the mass region A=50-65 have been studied experimentally via $(n,p)$ charge 
exchange reactions at forward angles 
\cite{vetterli,ronnqvist,kateb,williams,alford2}.
The $(n,p)$ charge-exchange reaction is one of the best efficient ways to 
extract the $GT_+$ strength in nuclei. For incident energies above 100 MeV, 
the isovector spin-flip component of the effective interaction is dominant 
and the  cross sections arise mainly from spin-isospin transitions.
At forward angles and low excitation energies in the final nucleus, the 
momentum transfer is small and therefore the reaction cross section is 
dominated by the GT operator with 
$\Delta T=1, \Delta L=0, \Delta J^\pi=1^+$. The cross section, 
extrapolated to zero momentum transfer, is proportional to the 
$\beta$-decay strength between the same states. Charge exchange 
reactions at small momentum transfer can therefore be used to study 
GT strength distributions when $\beta$-decay is not energetically 
possible.

The experimental data \cite{vetterli,ronnqvist,kateb,williams,alford2} show
that the total $GT_+$ strength is strongly quenched and fragmented over 
many final states, as compared to the independent particle model. This is 
caused by residual nucleon nucleon correlations. The data also indicate 
a systematic misplacement of the GT centroid adopted in the 
parameterizations of Ref. \cite{ffn}.
An improved theoretical description of the stellar weak interaction rates,
treating the nuclear structure problem more accurately, is then required.
Shell Model Montecarlo \cite{smmont} and large scale Shell Model 
diagonalization calculations \cite{sm1,sm2} have been already used to 
derive the stellar rates. The reliability of the latter diagonalization 
methods was demonstrated \cite{sm2} by comparing the calculated $GT_+$ 
strength distributions of nuclei in the iron mass region with the 
corresponding experimental distributions in the whole range of excitation 
energies, as obtained from $(n,p)$ charge exchange reactions.

This comparison with experiment is important not only for the direct
determination of the $GT_+$ strength distributions in stable nuclei, 
but also for the calibration of model calculations that are used in a
further step to estimate the strength distributions for unstable nuclei 
or the strength distributions of nuclei under high temperature and density
conditions, where no experimental information is available.
In this line of thought, it is also important to compare to the data set 
\cite{vetterli,ronnqvist,kateb,williams,alford2} the predictions of other 
microscopic models that, though may not be as accurate as the Shell Model 
calculations of Caurier et al. \cite{sm2} in this mass region ($A\sim 60$),
have at present a wider range of applicability. This is particularly the 
case of the proton-neutron quasiparticle random phase approximation (pnQRPA)
with separable Gamow-Teller ($V_{GT}$) residual interaction. The purpose of
this work is to test to what extent this approach, not limited by mass 
number, can account for the above mentioned set of data.

The pnQRPA method with a separable GT (or Fermi, $V_F$) interaction was 
first proposed and applied in Ref. \cite{sph}, on a spherical harmonic 
oscillator basis, and then it was extended to deformed nuclei 
\cite{moller} using deformed phenomenological single-particle basis. 
Like standard RPA, for a repulsive residual interaction, the method is 
correct both in the weak coupling and strong coupling limits. 
The strong coupling limit of pnQRPA gives the correct result by yielding 
the multiplet (supermultiplet) structure associated with the $V_F$ 
($V_{GT}$) force, respectively \cite{sph}. Other attractive features of 
the pnQRPA method are that the linear energy weighted sum rule is 
conserved and that Ikeda sum rule is fulfilled.
Further refinements to the pnQRPA formalism have been introduced along the 
years (see in particular \cite{refine,sarr1} and references therein), 
including, in particular, particle-particle residual interactions 
\cite{engel,klapdor,homma}.

Weak interaction rates and nuclear properties relevant for astrophysical 
applications within the pnQRPA have also been reported \cite{qrpacompil}, 
but a detailed comparison of the predictions of this model with 
experimental distributions in the iron mass region is still missing. In 
Refs. \cite{sarr1,sarr2,sarr3,sarr4} we performed pnQRPA calculations 
based on a deformed Hartree-Fock basis obtained from density dependent 
Skyrme forces and pairing correlations. In this work we extend those 
calculations to the iron mass region.

The paper is organized as follows: In the next Section we describe briefly
the theoretical formalism used to calculate the GT strength distributions.
In Section 3 we present and discuss the results obtained and compare them
with experiment. In Section 4 we summarize the main conclusions. 

\section{Theoretical formalism}

In a previous work \cite{sarr1,sarr2,sarr3,sarr4} we studied ground state 
and $\beta $-decay properties of even-even and odd-A exotic nuclei on the 
basis of a deformed selfconsistent HF+BCS+pnQRPA calculation with density 
dependent effective interactions of Skyrme type, including $T_z=\pm1$ 
pairing correlations in BCS approximation.
Our purpose here is to extend these calculations to stable nuclei in
the iron mass region and to investigate up to what extent this approach is 
able to reproduce the experimental information extracted from the charge 
exchange reactions in the Fe-Ni region.

The theory involved in the microscopic calculations can be seen in detail 
in Refs. \cite{sarr1,sarr2,sarr3,sarr4}.
For the solution of the Hartree-Fock (HF) equations we follow the McMaster 
procedure \cite{sprung} that is based in the formalism developed in 
Ref. \cite{vautherin}. 
Time reversal and axial symmetry are assumed. The single-particle wave 
functions are expanded in terms of the eigenstates of an axially symmetric 
harmonic oscillator in cylindrical coordinates using eleven major shells 
in the expansion. The method also includes pairing between like nucleons 
in the BCS approximation with fixed gap parameters for protons 
$\Delta _{\pi},$ and neutrons $\Delta _{\nu}$, which are determined 
phenomenologically from the odd-even mass differences through a symmetric 
five term formula involving the experimental binding energies \cite{audi}. 

For odd-A nuclei, the fields corresponding to the different interactions 
were obtained from the corresponding selfconsistent field of the closest 
even-even nucleus, selecting the orbital occupied by the odd nucleon among 
those around Fermi level, according to the experimental spin and parity.

We study first the energy surfaces as a function of deformation for all 
the isotopes under study here. For that purpose, we perform constrained
HF calculations with a quadrupole constraint \cite{constraint} and
minimize the HF energy under the constraint of keeping fixed the nuclear 
deformation. The GT distributions in the next section are then calculated 
for the equilibrium shape of each nucleus obtained in this way, that is, 
for the solution, in general deformed, for which we obtain the minimum 
in the energy surface.

We can see in Fig. 1 the total HF+BCS energy as a function of deformation
for the nuclei under study in this work with three different Skyrme 
interactions:
SG2 \cite{giai}, Sk3 \cite{sk3}, and the more recent SLy4 \cite{sly4}.
For an easier comparison, the origin is different in the vertical axis
for the three forces but the distance between ticks corresponds always to 
1 MeV. As we can see from the figure, the three forces predict a similar
behavior in most cases. Thus, SG2 and SLy4 produce a shallow minimum
around the spherical shape in $^{54}$Fe with a slightly prolate shape
favored energetically. On the other hand, Sk3 force gives a spherical
equilibrium shape. In the case of $^{56}$Fe, the three forces favor a
prolate solution with another oblate minimum at about 1 MeV higher.
$^{58}$Ni is spherical according to the results obtained from the three
forces considered. In the case of Ni isotopes, we observe that the SLy4
interaction predicts spherical shapes in $^{58}$Ni, $^{60}$Ni and
$^{62}$Ni, although the minimum is shallower in the latter cases. 
In $^{64}$Ni a slightly oblate shape is predicted with this force. 
The SG2 and Sk3 interactions produce again a spherical shape in $^{58}$Ni, 
but contrary to the force SLy4, a shape coexistence  between oblate and 
prolate shapes  is predicted in $^{60,62,64}$Ni.

We can also see in Table 1 a comparison of experimental and calculated
charge radii ($r_c$) and quadrupole moments ($Q_0$). Experimental values
are from Refs. \cite{devries} in the case of $r_c$ and from \cite{raghavan} 
in the case of $Q_0$. The calculated values are obtained from deformed 
HF+BCS calculations with the force SG2.

To describe GT transitions we add to the mean field a 
spin-isospin residual interaction. This interaction contains two parts, 
particle-hole ($ph$) and particle-particle ($pp$). The $ph$ part is 
mainly responsible for the position and structure of the GT resonance 
\cite{klapdor,sarr2} and, in principle, it could be derived 
selfconsistently from the same energy density functional as the HF 
equation. After averaging the force over the nuclear volume, it can be 
written in a separable form \cite{sarr1,sarr2}, with a coupling strength 
$\chi ^{ph}_{GT}$ determined by the Skyrme parameters. By taking 
separable GT forces, the energy eigenvalue problem 
reduces to find the roots of an algebraic equation.

Since the GT giant resonance for stable nuclei can be measured 
from charge exchange reactions, it is a common practice to fit the 
coupling strength $\chi ^{ph}_{GT}$ to reproduce the energy of the 
resonance. Several parameterizations have been proposed to reproduce 
these energies in large scale calculations. Nevertheless, one should 
take into account that the coupling strengths obtained in this way depend
on the model used for single particle and correlated wave functions and
on the set of experimental data considered. Thus, values of the 
coupling strengths obtained from a given fitting procedure cannot be
safely extrapolated to other cases.

The particle-particle part is a neutron-proton pairing force in the 
$J^\pi=1^+$ coupling channel. We introduce this interaction in the usual 
way \cite{engel,klapdor,sarr3}, that is, in terms of a separable force 
with a coupling constant $\kappa ^{pp}_{GT}$, which is usually fitted to 
reproduce the half-lives.

Both $ph$ and $pp$ residual interactions reduce the GT strength. 
Calculations within a single major shell in both QRPA and configuration 
mixing Shell Model were made in Ref. \cite{auerbach}, where it was shown 
that the quenching produced by QRPA calculations is similar to that found 
in Shell Model including all possible $0p0h$ and $2p2h$ configurations. 
The residual forces produce also a displacement of the GT strength, 
which is to higher energies in the case of the repulsive $ph$ force and 
to lower energies in the case of the attractive $pp$ force.

The optimum set of coupling strengths 
$(\chi ^{ph}_{GT}, \kappa ^{pp}_{GT})$ could be chosen following a case 
by case fitting procedure and we will get different answers depending on 
the nucleus, shape and Skyrme force. However, since the purpose here is 
to test the ability of pnQRPA models to account for the GT strength 
distributions in the iron mass region with as few free parameters as 
possible, we have chosen to use the same coupling strengths for all 
the nuclei considered in this work. 

The pnQRPA phonon operator for GT excitations in even-even nuclei is 
written as

\begin{equation}
\Gamma _{\omega _{K}}^{+}=\sum_{\pi\nu}\left[ X_{\pi\nu}^{\omega _{K}}\alpha
_{\nu}^{+}\alpha _{\bar{\pi}}^{+}-Y_{\pi\nu}^{\omega _{K}}
\alpha _{\bar{\nu}}\alpha_{\pi}\right]\, ,  
\label{phon}
\end{equation}
where $\pi$ and $\nu$ stand for proton and neutron, respectively,
$\alpha ^{+}\left( \alpha \right) $ are quasiparticle creation
(annihilation) operators, $\omega _{K}$ are the RPA excitation energies, 
and $X_{\pi\nu}^{\omega _{K}},Y_{\pi\nu}^{\omega _{K}}$ the forward and 
backward amplitudes, respectively. It satisfies
\begin{equation}
\Gamma _{\omega _{K}} \left| 0\right\rangle=0\, ; \qquad
\Gamma ^+ _{\omega _{K}} \left| 0\right\rangle = \left|
\omega _K \right\rangle .
\end{equation}
Solving the pnQRPA equations \cite{moller,sarr1,klapdor},
the GT transition amplitudes in the intrinsic frame of even-even nuclei
connecting the QRPA ground state of the parent nucleus
$\left| 0\right\rangle $ to one phonon states in the daughter nucleus 
$\left| \omega _K \right\rangle $, 
are given by
\begin{equation}
\left\langle \omega _K | \beta _K^{\pm} | 0 \right\rangle = \mp 
M^{\omega _K}_\pm \, ,
\label{ampleven}
\end{equation}
where  $\beta _K^{\pm}=\sigma_K \tau^{\pm},\quad K=0,\pm 1$, and 

\begin{equation}
M_{-}^{\omega _{K}}=\sum_{\pi\nu}\left( q_{\pi\nu}X_{\pi\nu}^{\omega _{K}}+
\tilde{q}_{\pi\nu}Y_{\pi\nu}^{\omega _{K}}\right) \, ; \qquad 
M_{+}^{\omega _{K}}=\sum_{\pi\nu}\left( \tilde{q}_{\pi\nu}
X_{\pi\nu}^{\omega _{K}}+
q_{\pi\nu}Y_{\pi\nu}^{\omega _{K}}\right) \, ,
\end{equation}
with
\begin{equation}
\tilde{q}_{\pi\nu}=u_{\nu}v_{\pi}\Sigma _{K}^{\nu\pi };\ \ \ 
q_{\pi\nu}=v_{\nu}u_{\pi}\Sigma _{K}^{\nu\pi}\, ;\qquad
\Sigma _{K}^{\nu\pi}=\left\langle \nu\left| \sigma _{K}\right| 
\pi\right\rangle \, ,
\label{qs}
\end{equation}
where $v'$s are occupation amplitudes ($u^2=1-v^2$).
The single particle wave functions, energies, and occupation probabilities 
are generated from the selfconsistent deformed mean field obtained with the 
Skyrme force.
To calculate GT strengths we have considered the force SG2
that has been successfully tested against spin-isospin excitations in 
spherical \cite{giai} and deformed nuclei 
\cite{sarr1,sarr2,sarr3,sarr4,sarrnoj}.
Comparison to calculations obtained with other Skyrme forces have been made
in Refs. \cite{sarr1,sarr2}, showing that the results do not differ in a 
significant way.

When the parent nucleus has an odd nucleon, the
ground state can be expressed as a one quasi-particle state in which the 
odd nucleon occupies the single-particle orbit of lowest energy. 
Then two types of transitions are possible. One type is due to phonon
excitations in which the odd nucleon acts only as a spectator. In the 
intrinsic frame, the transition amplitudes are in this case basically 
the same as in the even-even case  but with the blocked spectator 
excluded from the calculation.
The other type of transitions are those involving the odd nucleon state,
which are treated \cite{moller,klapdor,sarr4} by taking into account phonon 
correlations in the quasiparticle transitions in first order perturbation.

Once the intrinsic amplitudes 
$\left< f \left|  \beta ^\pm _{K} \right| i\right>$
are calculated, 
the Gamow-Teller strength $B_{GT}$ for a transition $I_i \rightarrow I_f$
can be obtained as

\begin{eqnarray}
B^{\pm}_{GT}&=&\sum_{M_i,M_f,\mu} \left| 
\left< I_fM_f \left| \beta ^\pm _\mu
\right| I_i M_i \right> \right|^2 \nonumber \\
&=& \delta_{K_i,K_f} \left[ \left< I_i K_i 1 0 | I_f K_f \right> \left< 
\phi_{K_f} \left|  \beta ^\pm _0 \right| \phi_{K_i}\right> \right.
\nonumber \\ 
&& \left. +\delta_{K_i,1/2}(-1)^{I_i-K_i} \left< I_i -K_i 1 1 | I_f K_f \right> 
\left< \phi_{K_f} \left|  \beta ^\pm _{+1} \right| \phi_{\bar{K_i}}\right> 
\right] ^2 \nonumber \\
&& +\delta_{K_f,K_i+1}  \left< I_i K_i 1 1 | I_f K_f \right> ^2 \left< 
\phi_{K_f} \left|  \beta ^\pm _{+1} \right| \phi_{K_i}\right> ^2\nonumber \\
&&+\delta_{K_f,K_i-1} \left< I_i K_i 1 -1 | I_f K_f \right> ^2 \left< 
\phi_{K_f} \left|  \beta ^\pm _{-1} \right| \phi_{K_i}\right> ^2 \, ,
\label{strodd}
\end{eqnarray}
in units of $g_A^2/4\pi$.
To obtain this expression we have used the initial and final states in the 
laboratory frame expressed in terms of the intrinsic states 
$\left| \phi_K\right>$ using the Bohr-Mottelson factorization \cite{bm}.

Eq. (\ref{strodd}) can be particularized for even-even parent nuclei. 
In this case $I_i=K_i=0$, $I_f=1$, and $K_f=0,1$.

\begin{equation}
B^{\pm}_{GT}=\frac{g_A^2}{4\pi}\left\{ \delta_{K_f,0}
\left< \phi_{K_f} \left|  \beta ^\pm _0 \right| \phi_0\right> ^2
+2\delta_{K_f,1}
\left< \phi_{K_f} \left|  \beta ^\pm _1 \right| \phi_0\right> ^2 \right\} 
\, .
\label{streven}
\end{equation}

\section{Results}

In this Section we present and discuss the results obtained for the GT 
strength distributions and summed strengths. 
The results correspond to pnQRPA calculations with the Skyrme force SG2 and 
they have been performed for the nuclear shape that minimizes the HF 
energy. 
Before discussing the figures we note that the GT strength distributions 
are plotted versus the excitation energy of the daughter nucleus.
The distributions of the GT strength  have been folded with $\Gamma =2$
MeV width Gaussians to account for the finite experimental resolution as
it was done in Refs. \cite{smmont,sm1,sm2}, 
so that the original discrete spectrum is transformed into a continuous 
profile.
The theoretical GT distributions in those figures have been quenched 
with a factor
$[(g_A/g_V)_{\mbox{\scriptsize{eff}}}/
(g_A/g_V)_{\mbox{\scriptsize{free}}}]^2 =(0.7)^2$,
which is standard for transitions involving the spin operator \cite{towner}.
The observed GT strength in charge exchange reactions is less than the 
expected strength from the Ikeda sum rule.

This quenching factor is similar to that found in spin $M1$ transitions 
in stable nuclei, where 
$g_{s,{\mbox{\scriptsize{eff}}}}$ is also known to be approximately 
$0.7\ g_{s,{\mbox{\scriptsize{free}}}}$. 

As we have already mentioned, the two coupling strengths of the $ph$ and
$pp$ residual interactions have been determined to reproduce the positions
of the experimental $GT_+$ resonances as obtained from the $(n,p)$
reactions. This has been done in a global way, choosing 
$\chi ^{ph}_{GT}=0.10$ MeV and $\kappa ^{pp}_{GT}=0.05$ MeV, as the best 
set of values within a deformed HF basis with the force SG2 for  
the observed resonances in the set of nuclei considered in 
this work. We note that the value of $\chi ^{ph}_{GT}$, which mainly 
determines the position of the resonance, is smaller than the typical 
values expected from systematic fits \cite{homma} or from consistency 
with the SG2 mean field \cite{sarr1,sarr2,sarr3,sarr4}. This feature was 
already pointed out in Ref. \cite{hirsch}, where the coupling strengths 
$\chi ^{ph}_{GT}$ needed to reproduce the GT giant resonances were 
found to be smaller than the fitted $A^{-1}$ law, in the mass region 
under consideration in this work.

The results obtained for the $GT_+$ strength distributions 
can be seen in Fig.2. Plotted downward are the results of uncorrelated 
two-quasiparticle (HF+BCS) calculations, both individual and folded 
strengths. Plotted upward are the experimental data, accumulated in 
1 MeV bins (dots with error bars), as well as the $GT_+$ strengths 
obtained from correlated two-quasiparticle (HF+BCS+ pnQRPA) calculations. 
The agreement with experiment is in general quite satisfactory. We can 
see that the experimental $GT_+$ strength distributions  are fragmented 
over many states and that the centroids and widths of the distributions 
are well reproduced in our calculations. However, experimental $GT_+$ 
distributions show a tendency to build up a second peak beyond 
$\sim6$ MeV, which is not seen in the theoretical
calculations. This indicates that in order to reproduce these strengths at
high energy, one should include higher correlations in the theoretical 
calculations. The total $GT_+$ strength contained below the measured 
excitation energies can be seen in Table 2. We can see that the summed 
strengths in HF+BCS+pnQRPA  agree better with experiment, but are still 
somewhat larger than the experimental ones.

We observe that the inclusion of the pnQRPA correlations
reduces the total HF+BCS strength by about 20-40\%, depending on the 
nucleus, improving always the comparison with
experiment. However, even the simpler HF+BCS calculation with the force SG2
produces quite good results in most cases. This is not only true for the
$GT_+$ strength distributions, but also for the summed strengths.
Our theoretical summed strengths are in general somewhat larger than 
those from Shell Model calculations \cite{smmont,sm2}, however, in 
some instances compare better to experiment. This is the case of 
$^{64}$Ni, where Shell Model calculations \cite{smmont,sm2} give 
strengths considerably lower than experiment.

Though the $GT_+$ strength distributions obtained from full Shell Model 
calculations by Caurier et al. \cite{sm2} may agree better with experiment 
in some details, the present HF+BCS+pnQRPA results are, on the overall, 
of comparable quality. The problem of missing theoretical strength at 
high energy (that was found in Shell Model calculations) persist here, 
although we get more strength at higher energy than Shell Model 
\cite{sm2} because of the higher N-shell mixing contained in HF+BCS+pnQRPA.

The peaks of the observed $GT_+$ strength distributions in the odd-A
nuclei considered are found to be consistently at higher excitation
energies in the daughter nucleus as compared to their even-even partners.
This feature is well described in our calculations and the reason, which 
was discussed in Ref. \cite{sarr4}, is related to the energy needed to
break a Cooper pair. 

Although the stellar electron captures and $\beta$-decays are more 
sensitive to the distribution of the $GT_+$ strength, the $GT_-$ strength 
distributions play also a non negligible role in the 
calculation of the stellar weak interaction rates and
it is also of interest the study of their distributions. 
In addition, this allows to study Ikeda sum rule, which
is always fulfilled in our calculations, as well as the total quenching.

We plot in Fig. 3  the experimental forward-angle $(p,n)$ cross sections  
from Ref. \cite{rapa} because only in a few specific cases \cite{anderson}
the cross sections have been converted into $GT_-$ strength distributions.
We can observe the highly fragmented strength distribution obtained as
well as the concentration of the strength in different energy regions
depending on the nucleus. In the lower panel we can see the calculated 
$GT_-$ strength distributions obtained from HF+BCS+pnQRPA calculations with
the force SG2. We show three different types of results for each nucleus.
The dotted lines are the HF+BCS results without including any 
residual interaction. The curves labeled $QRPA_1$ and $QRPA_2$ are the
results obtained when we introduce the $ph$ and $pp$ residual interactions
discussed above with two different sets of parameters for  $\chi ^{ph}_{GT}$
and  $\kappa ^{pp}_{GT}$. The dashed lines ($QRPA_1$) are obtained using
the same coupling strengths we have used to calculate $GT_+$ strengths in
Fig. 2, that is  $\chi ^{ph}_{GT}=0.10$ MeV and  $\kappa ^{pp}_{GT}=0.05$ 
MeV. For comparison we also show by solid lines ($QRPA_2$) results obtained 
using the coupling strengths derived from the Skyrme force following the 
procedure in Ref. \cite{sarr1} ($\chi ^{ph}_{GT}$=0.43-0.55 MeV, 
depending on the nucleus), and using $\kappa ^{pp}_{GT}$=0.07 MeV. 
These values are close to the values derived from the 
parameterization of Ref. \cite{homma} 
($\chi ^{ph}_{GT}=5.2/A^{0.7}$=0.28-0.34 MeV, 
$\kappa ^{pp}_{GT}=0.58/A^{0.7}$=0.032-0.038 MeV, depending also on the 
nucleus). Though here we cannot compare directly to $GT_-$ data, and 
Fermi contributions are also included, the figure shows that theory 
reproduces the two peak structure, which is the dominant feature of data. 
Clearly, with the same set of parameters we also get missing $GT_-$ 
strength in the high energy sector as it was the case in $GT_+$.

\vfill\eject

\section{Conclusions and final remarks}

We have applied a selfconsistent deformed HF+BCS+pnQRPA formalism with
density-dependent effective Skyrme interactions to the description of 
the GT strength in several nuclei in the Fe-Ni mass region. 

We find that the present pnQRPA calculations are able to reproduce the
main features of the GT properties measured in this mass region, 
reinforcing confidence in the method and in its predictive power. The
method had been successfully contrasted against the experimental half-lives
of unstable proton rich nuclei in the $A\sim 70$ mass region on one hand,
and against experimental $M1$ spin strength distributions in the rare earths
region on the other. In those applications we found agreement with 
experiment using the Gamow-Teller strength constant ($\chi ^{ph}_{GT}$) 
derived consistently with the mean field from the same Skyrme interaction, 
following a procedure  that gives an overall $1/A$ law for 
$\chi ^{ph}_{GT}$ \cite{sarr1}.
From phenomenological fits \cite{hirsch} it is known that the empirical 
$1/A$ law does not work well in this mass region, where fitted 
$\chi ^{ph}_{GT}$ values are considerably lower than those suggested by 
the $1/A$ line. It is therefore not surprising that we find here better 
agreement with data with a smaller $\chi ^{ph}_{GT}$. The reasons why 
the overall procedure used in \cite{sarr1}, or equivalently the $1/A$ 
laws do not apply here is an interesting subject for future investigation 
in itself. It may be connected to the fact that those methods do not 
take into account possible renormalization of the  GT strength
constant in the vicinity of shell closures. It may also be connected 
to particular properties of effective two-body forces of Skyrme type 
that have so far not been sufficiently investigated. Clearly, the fact 
that $\chi ^{ph}_{GT}$ is small in this region implies that pnQRPA 
correlations are smaller than in other regions previously investigated, 
and that the bare two quasiparticle approximation is already a fairly 
good approximation here. In the latter, only deformation and 
$T_z=\pm1$ pairing correlations are taken into account. We find that 
deformation is essential in these calculations to get the proper 
fragmentation of the strength, that clearly shows up and differentiates
the experimental strength distributions of the different nuclei.

Although in this mass region the Shell Model calculations of Ref. \cite{sm2}
may be superior, and some fine details of data may be better described, 
we find that on the overall the agreement with experiment of present 
pnQRPA calculations is comparable and tend to do better in the higher 
energy domain ($E> 7$ MeV). However, in this domain both methods fail 
to reproduce data. Provided these data are unquestionable, they point 
out to a lack of higher order correlations (as it is particularly for 
pnQRPA) or to limitations of the single particle basis (as it is 
particularly for Shell Model) in the theoretical calculations. With 
these limitations in mind, the present comparison of pnQRPA to data 
provides a fairly sound basis to safely apply this method to the 
estimates of GT strengths and particularly of $\beta-$decay properties 
of highly unstable nuclei in other mass regions. In addition, our 
approach can be extended to much heavier nuclei beyond the present 
capability of the full Shell Model, without increasing the complexity 
of the calculations. 

\acknowledgments 

This work was supported by Ministerio de Ciencia y Tecnolog\'{\i}a (Spain) 
under contract number BFM2002-03562.  One of us (R.A.-R.) thanks Ministerio
de Educaci\'on, Cultura y Deporte (Spain) for financial support.

\vfill\eject

\vfill \eject

\begin{table}[tbp]
\begin{center}

{\bf Table 1.} Comparison of the experimental charge radii [fm] 
\cite{devries} and quadrupole moments [b] \cite{raghavan} with the 
calculated values obtained with the force SG2.

\vskip 0.5cm
\begin{tabular}{lcccc}
%\hline \hline
& $r_c$ exp & $r_c$ calc & $Q_0$ exp & $Q_0$ calc \\ 
\\
% $^{51}$V  & 3.58 -- 3.62 & 3.61 & -0.15(4) & -0.05\\
$^{54}$Fe & 3.675 -- 3.732 & 3.72 & +0.18(49) & +0.46\\
% $^{55}$Mn & 3.68 & 3.74 & +1.50(5) & +0.74 \\
$^{56}$Fe & 3.721 -- 3.801 & 3.74 & +0.81(10) & +0.74 \\
$^{58}$Ni & 3.769 -- 3.772 & 3.81 & +0.35(21) & +0.40 \\
% $^{59}$Co & 3.78 -- 3.80 & 3.80 & +0.86(9) & +0.81 \\
$^{60}$Ni & 3.793 -- 3.797 & 3.84 & -0.11(17) & -0.64 \\
$^{62}$Ni & 3.822 -- 3.830 & 3.87 & -0.18(42) & -0.80 \\
$^{64}$Ni & 3.845 -- 3.907 & 3.88 & -1.4(7) & -0.60 \\
%\hline \hline
\end{tabular}
\end{center}
\end{table}

\vspace {2cm}

\begin{table}[tbp]
\begin{center}
{\bf Table 2.} Comparison of the $GT_+$ strengths in the energy range 
experimentally available ($E_{ex}$) between experimental
\cite{vetterli,ronnqvist,kateb,williams,alford2}
measurements and theoretical calculations.
\vskip 0.5cm
\begin{tabular}{lccc}
%\hline \hline
& exp & HF+BCS & HF+BCS+pnQRPA \\ 
\\
$^{51}$V  & 1.2 $\pm$ 0.1 ($E_{ex} \leq $ 8 MeV) & 2.03 & 1.61 \\
$^{54}$Fe & 3.5 $\pm$ 0.7 ($E_{ex} \leq $ 9 MeV) & 5.14 & 4.24 \\
$^{55}$Mn & 1.7 $\pm$ 0.2 ($E_{ex} \leq $ 8.5 MeV)  & 2.72 & 2.18 \\
$^{56}$Fe & 2.9 $\pm$ 0.3 ($E_{ex} \leq $ 8.5 MeV)  & 4.16 & 3.24 \\
$^{58}$Ni & 3.8 $\pm$ 0.4 ($E_{ex} \leq $ 8.5 MeV)  & 6.19 & 5.00 \\
$^{59}$Co & 1.9 $\pm$ 0.1 ($E_{ex} \leq $ 8 MeV) & 3.26 & 2.50 \\
$^{60}$Ni & 3.11 $\pm$ 0.08 ($E_{ex} \leq $ 8 MeV)  & 4.97 & 3.72 \\
$^{62}$Ni & 2.53 $\pm$ 0.07 ($E_{ex} \leq $ 8 MeV)  & 3.40 & 2.36 \\
$^{64}$Ni & 1.72 $\pm$ 0.09 ($E_{ex} \leq $ 8 MeV)  & 2.65 & 1.65 \\ 
%\hline \hline
\end{tabular}
\end{center}

\end{table}

\newpage

\begin{figure}[t]
\begin{center}

\epsfig{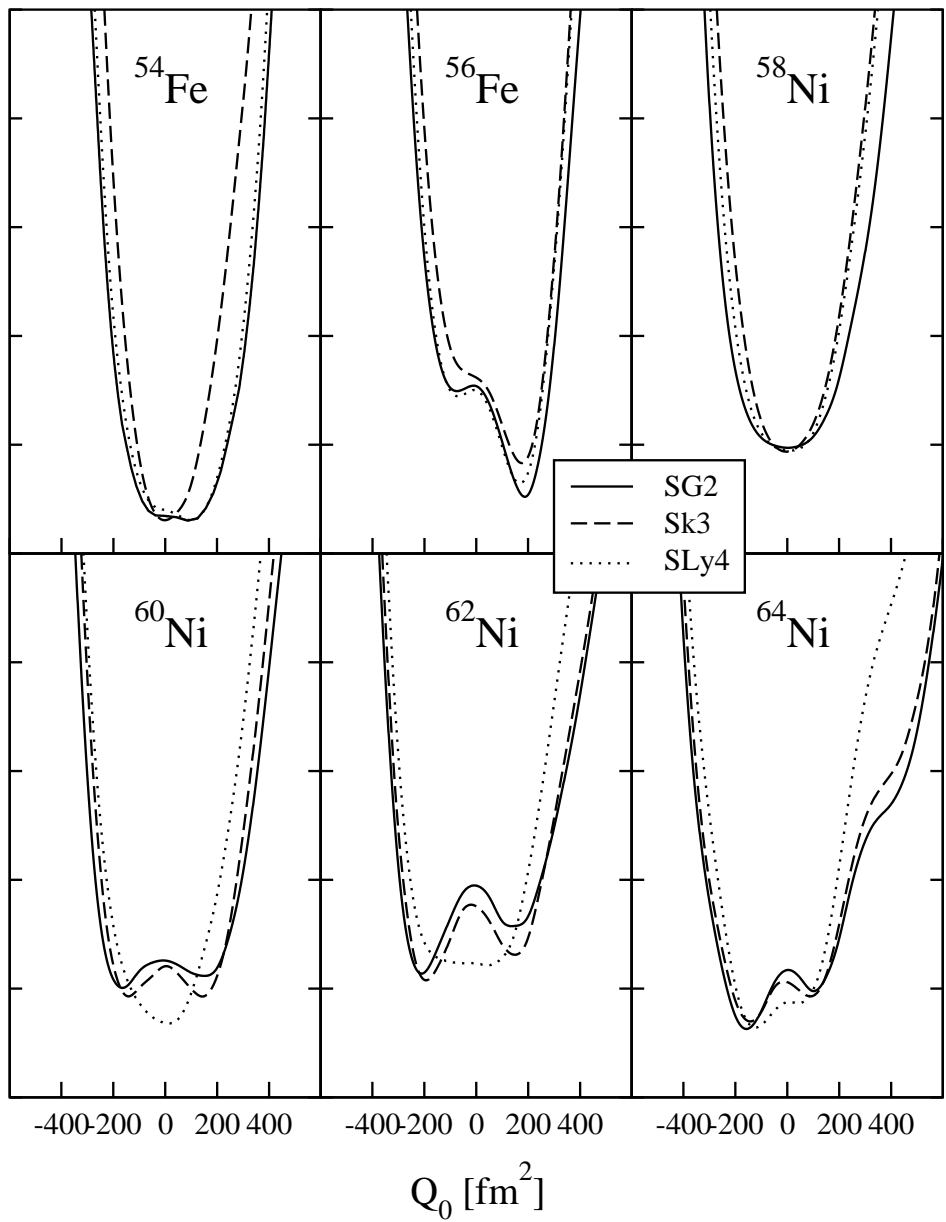}
\vskip 1cm
\caption{ Total energy as a function of the mass quadrupole moment 
$Q_0$ obtained from a constraint HF+BCS calculation with three Skyrme 
forces SG2, Sk3, and SLy4. }
\end{center}
\end{figure}

\vfill\eject

\begin{figure}[t]
\begin{center}

\epsfig{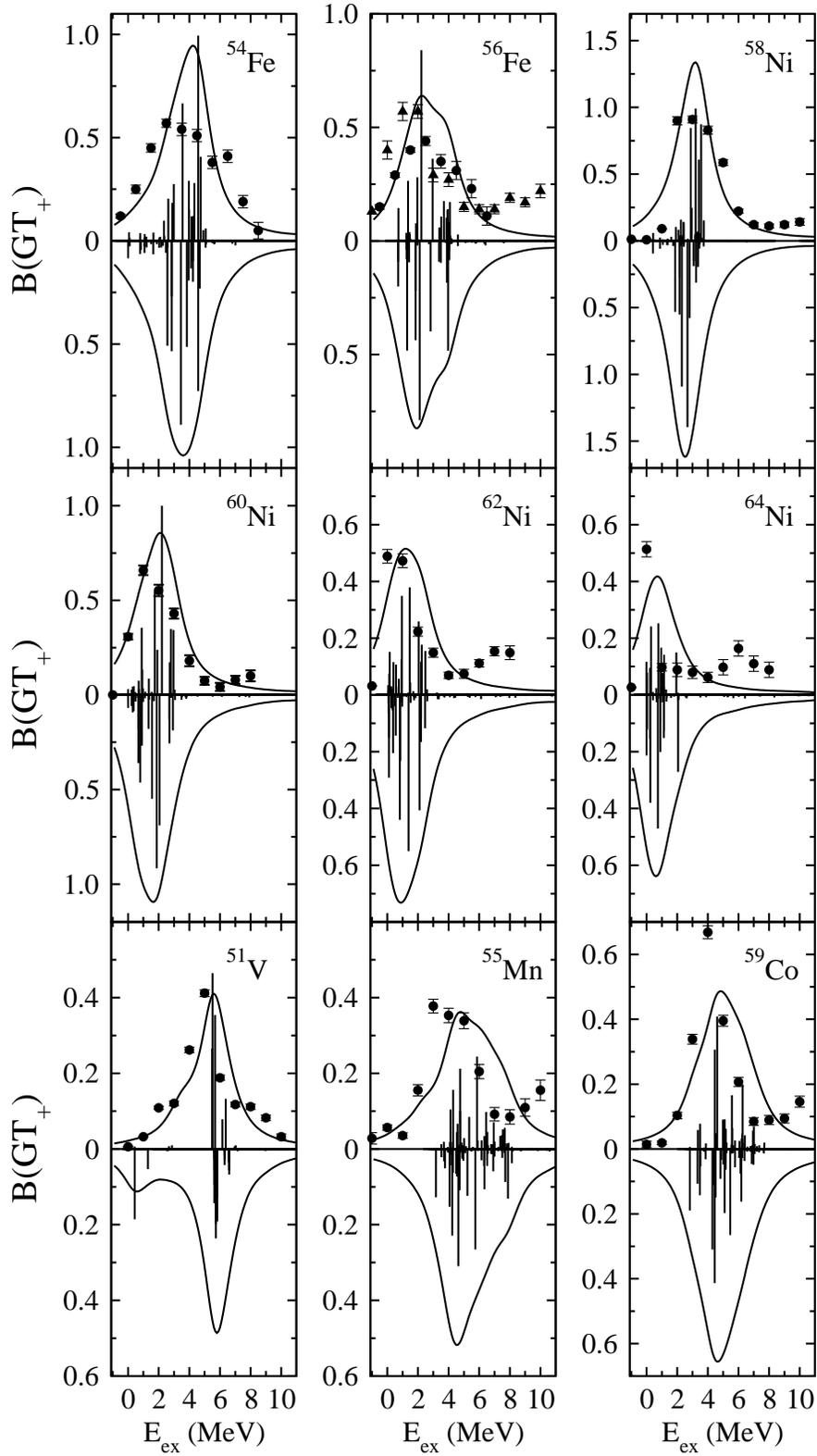}
\vskip 1cm
\caption{ Gamow-Teller strength distributions ($GT_+$) plotted versus 
the excitation energy of the corresponding daughter nucleus. The results 
are plotted downward in the case of HF+BCS approximation and 
upward in the case of HF+BCS+pnQRPA. Experimental data are from 
Refs. [3-7].}
\end{center}
\end{figure}

\vfill\eject

\begin{figure}[t]
\begin{center}
\epsfig{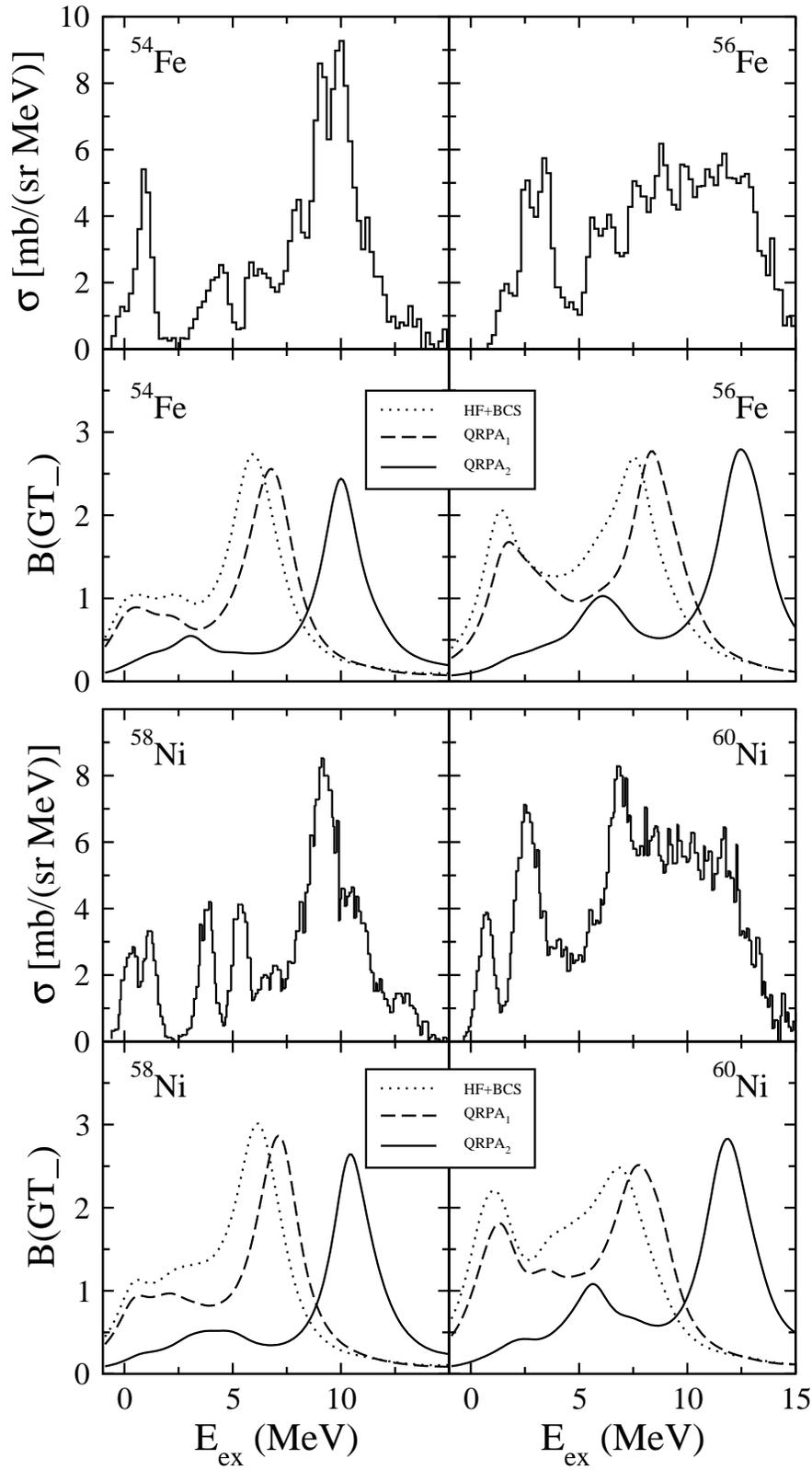}
\vskip 1cm
\caption{ Comparison of $(p,n)$ $L=0$ forward-angle cross section data
[37] (upper panels) with the calculated Gamow-Teller strength 
distributions ($GT_-$) as a function of the excitation energy of the 
daughter nucleus [MeV]. }
\end{center}
\end{figure}

\vfill\eject

\end{document}